\documentclass[aps,pre,reprint,showpacs]{revtex4-1}
\usepackage{amssymb,graphicx}

\def\o{\mathbf{\omega}}

\def\l{{\cal L}}

\def\T{\mathbf{Tr}}

\def\id{{\cal D}}

\def\l{{\cal L}}

\def\z{{\cal Z}}
\def\s{{\cal S}}

\def\pd{\partial}
\def\d{\mathrm{d}}

\def\exp{\mathrm{exp}}
\def\ex{\mathrm{e}}
\def\be{\begin{equation}}
\def\ee{\end{equation}}
\def\bea{\begin{eqnarray}}
\def\eea{\end{eqnarray}}
\def\ie{\textit{i.e.} }
\def\etal{\textit{et.al.} }

\begin{document}

\title{Anharmonic oscillation effect on the Davydov-Scott monomer in thermal bath }

\author{A. Sulaiman$^{1,2,4}$}
\email[Email:]{asulaiman@webmail.bppt.go.id,sulaiman@teori.fisika.lipi.go.id}
\author{F.P. Zen$^{1,4}$}
\email[Email:]{fpzen@fi.itb.ac.id}
\author{H. Alatas$^{3,4}$}
\email[Email:]{alatas@ipb.ac.id}
\author{L.T. Handoko$^{5,6}$}
\email[Email:]{handoko@teori.fisika.lipi.go.id, handoko@fisika.ui.ac.id, laksana.tri.handoko@lipi.go.id}
\affiliation{$^1$Theoretical
Physics Laboratory, THEPI, Faculty of Mathematics and Natural
Sciences, Institut Teknologi Bandung, Jl. Ganesha 10 Bandung
40135, Indonesia}
\affiliation{$^2$ Badan Pengkajian dan Penerapan Teknologi, BPPT Bld. II (19$^{\rm th}$ floor), Jl. M.H. Thamrin 8, Jakarta 10340, Indonesia}
\affiliation{$^3$Theoretical Physics Division,  Department of Physics, Bogor Agricultural University, Jl. Meranti, Kampus IPB Darmaga, Bogor 16680, Indonesia}
\affiliation{$^4$Indonesia Center for Theoretical and Mathematical Physics, Jl. Ganesha 10, Bandung 40132, Indonesia}
\affiliation{$^5$Group for Theoretical and Computational Physics,
Research Center for Physics, Indonesian Institute of Sciences,
Kompleks Puspiptek Serpong, Tangerang, Indonesia}
\affiliation{$^6$Department of Physics, University of Indonesia,
Kampus UI Depok, Depok 16424, Indonesia}

\begin{abstract}
The dynamics of Davydov-Scott monomer in a thermal bath with higher order
amide-site's displacement leads to anharmonic oscillation effect is investigated
using full-quantum approach and the Lindblad formulation of master equation. The
specific heat is calculated based on the thermodynamic partition function using
the path integral method. The temperature dependence of the specific heat is
studied. In the model the specific heat anomaly as pointed out in recent works
by Ingold \etal is also
observed. However it is found that the anomaly occurs at high temperature
region, and the anharmonic oscillation restores the positivity of specific heat.
\end{abstract}


\maketitle

\section{Introduction}
\label{sec:intro}

The study of molecular biophysics began with Fr$\ddot{\mathrm{o}}$lich's
hypotheses which assumes that an excitation from an atomic vibration related to
biological activity \cite{takeno}. Based on this idea, Davydov has developed a
quantum theory of protein to understand the mechanisms of energy transport in
molecular protein, in particular alpha helix protein.  The mechanism is an
excitation energy of an amide-I is stabilized by its vibration in a combined
excitation which propagate as a soliton \cite{scott,scott2}. Most of studies in
this field have been done at zero temperature, and little attention has been
given at physiological  temperatures.

The studies of Davydov soliton in physiological temperature is realized by the Davydov soliton interacting with thermal bath. While zero temperature calculation allows the existence of a soliton states in the proteins, there is an important question whether these states are stable at biological temperatures \cite{cruzeiro1,cruzeiro2}. Because the temperature effect may exchange heat energy with surrounding aqueous medium. The measurement of infrared absorption and Raman's scattering  of an crystallineacetanilide $(CH_3CONHC_6H_5)_x$ at low temperature
showed a new band closing to the amide-I band \cite{careri1,careri2}. The result is interpreted as a signature of Davydov's soliton. The experiment using femtosecond IR spectroscopy realizing a band of the amide-I from acetanilide (ACN) and N-methylacetanide (NMA) shows the dependencies of absorption spectrum on temperature. At high temperature, the absorption spectrum shifts to higher frequency \cite{edler2}. Theoretical prediction using first order perturbation methods based on Davydov model gives soliton life time at the order of $O(10^{-12})$ s. This is very short time for biological processes in room temperature \cite{cottingham}.

Using standard Davydov model, some numerical calculations showed that soliton is stable at 310K \cite{cruzeiro1}. The calculation based on trial function by Kapor et.al. showed that soliton is stable at 300K \cite{kapor1}. The result has also been confirmed in \cite{kapor2}.

From those results, it is important to study Davydov model involving the contact
with thermal bath. The behavior of the system has attracted many interests in
the last three decades. The interaction of a system with its environment is
given by the dissipation effect in quantum system \cite{weiss,rigo,perceval}.
However, the dissipation effect leads to a serious problem for quantization
procedure due to the broken Heisenberg's relation. The most appropriate theory
to resolve this problem is the Linblad formulation of master equation
\cite{rigo,perceval}.

The first application of Linblad formulation of master equation to the protein
model has been done by Cuevas et.al.
\cite{cuevas}. They used Davydov-Scott monomer model and showed that at 10K the
quantum effect of amide-I vibration can not be neglected. Also at room
temperature the
semi classical approach might be a good approximation compared to the
corresponding full quantum system. However the study was focused on the
dynamical aspect of the system. Since the real world is affected by thermal
fluctuation, it is important to study such system from statistical mechanics
point of view.

In this paper the thermodynamic properties of Davydov-Scott
monomer is investigated using Lindblad formulation and the
partition function is calculated using path integral method
\cite{feynmann}. The path integral method is a powerful tool to
investigate the properties of nonlinear dynamical systems with
retarded interactions \cite{mazoli}. In Sec. \ref{sec:hdm} the
Hamiltonian of the system under consideration is described as a
coupled harmonics oscillation \ie  the amide-I and
amide-site oscillators. In addition to the original Davydov-Scott monomer, the
higher order of amide-site displacement inducing the anharmonic oscillation
effect of amide-site is also taken into account. This is motivated by the fact
that the excitation and relaxation of collective modes of a protein are
generally achieved via anharmonic interactions with other normal modes through
energy exchange \cite{xie}. In Sec. \ref{sec:pi} the partition function is
calculated using path integral approach to obtain the thermodynamic specific
heat of the system which is then discussed in Sec. \ref{sec:tpdc}.

\section{Hamiltonian of Davydov-Scott monomer with anharmonic oscillation effect
in Lindblad formulation of master equation}
\label{sec:hdm}

Considering the Davydov-Scott monomer, the excitation of the amide-I is
described by the coordinate ($x$) and momentum ($p$) operators. On the other
hand, the displacement and momentum operators of amide-site are expressed by $Q$
and $P$. The  hamiltonian for Davydov-Scott monomer can then be written in the
full quantum approach as
follow,
\be
   H = \frac{p^2}{2 m} + \frac{1}{2} m \o^2 x^2 - \frac{1}{4} \delta x^4 +
\frac{P^2}{2M} + V(Q)+ \chi \frac{\pd V}{\pd Q} x \; ,
   \label{eq:mono1}
\ee
where $\o$ is the intrinsic frequency of amide-I oscilation, $\chi$ measures
the coupling between the amide-I excitation and the amide site vibration, $m$
($M$) is the amide-I (amide-site) mass and $\delta$ is the anharmonic
coefficient. Here the amide-site potential can generally be expanded around the
origin,
\bea
  V(Q) & = & V_0 + \left. \frac{\pd V}{\pd Q} \right|_{Q=0} Q + \left.
\frac{1}{2}\frac{\pd^2 V}{\pd Q^2} \right|_{Q=0} Q^2
  \nonumber\\
  & & + \left. \frac{1}{6}\frac{\pd^3 V}{\pd Q^3} \right|_{Q=0} Q^3
      + \left. \frac{1}{24}\frac{\pd^4 V}{\pd Q^4} \right|_{Q=0} Q^4 + \cdots \;
.
  \label{eq:poten1}
\eea
The first term in Eq. (\ref{eq:poten1}) is a constant and can be scaled out. On
the other hand, the equilibrium condition implies that the term of
$Q$ vanishes due to $-\nabla V(Q)|_{Q=0} = 0$. The third term is the harmonic
oscillator one, while the remaining higher order terms represent the anharmonic
contributions.

The original Davydov-Scott model assumes that the anharmonicity of amide-site is
not important compared to the amide-I \cite{pouthier}. The nonlinear effects
appear only through exciton and amide-site coupling and higher order potential
of excitons. However, recent experiment in biopolymer showed that the
anharmonicity of polymer amide-site might be important  \cite{go,xie,petr}. The
solid state experiments showed that
 at high temperature (room temperature), specific heat is not a constant
anymore, but it is increasing. This fact can only be explained by taking into
account the anharmonic oscillator term  \cite{kittel}.

In Eq. (\ref{eq:poten1}), the terms $Q^3, Q^5, \cdots$ represent
the asymmetry of atom mutual repulsion, while the terms $Q^4, Q^6,
\cdots$ represent the softening of vibration at large amplitudes
\cite{kittel}. Moreover, concerning the fact that  the Davydov-Scott polymer has
an identic monomer, it is plausible to assume that the asymmetry of atom
mutual repulsion can be ignored. This yields,
\be
  V(Q) = \frac{1}{2} \, \kappa Q^2  - \frac{1}{4} \lambda \, Q^4  \; ,
  \label{eq:poten2}
\ee
where $\kappa/2 = 1/2 \, ({\pd^2 V}/{\pd Q^2})|_{Q=0}$ and $1/{24} \, ({\pd^4
V}/{\pd Q^4})|_{Q=0} = -\lambda/4$. Again, the second term induces the
anharmonic oscillation effect in the system.

Unfortunately the term leads to a severe problem of describing
the damping in open quantum systems, and it has been discussed for a long time.
One of the known models dealing with this problem is the one dimensional damped
harmonic oscillator (known as the Caldirola-Kanai Hamiltonian) \cite{um}. In
this model the momentum and coordinate operators are multiplied by $e^{-\gamma
t}$ and   $e^{\gamma t}$, where $\gamma$ is a damping factor.

In statistical mechanics, the behavior of an open system within system-plus-bath
can be modeled by the density matrix formalism $\rho$. The equation of density
matrix with hamiltonian $H$ and environment operator $R$ satisfy particular
master equation \cite{weiss}. It is usually restricted to weak system-bath
interaction. The density operator gives the probability for the expected
outcomes of measurements on the system. However, this formulation does not
preserve density operator properties, that is hermiticity, unit trace, and
positivity. The open quantum system theory which preserves density matrix
properties can be realized using Lindblad formulation. In this theory, the
interaction between Hamiltonian and thermal bath is realized by introducing some
operators called Lindblad operators. The  operators obey the master equation
\cite{lindblad},
\be
 \frac{\pd \rho}{\pd t} =
 -\frac{i}{\hbar}[H,\rho]+\frac{i}{2\hbar}
  \sum_{j} \left( [L_j, \rho L_j^\dag] + [L_j, \rho L_j^\dag] \right)  \; ,\
 \label{eq:lindblad}
\ee
where $L_j$ are the Lindblad operators. This choice is not unique and not
necessarily Hermitian. Since $L$ must be the first order in $Q$ and $P$
\cite{perceval}. The operators $H$ and $L$ denote the internal dynamics and
environmental effects of the system. Throughout the paper, the Lindblad
operators are put,
\bea
  L_1 &=& \sqrt{\gamma(1+\nu)}
  \nonumber \\
  && \times \left( \sqrt{\frac{M \Omega}{2\hbar}}Q+i\sqrt{\frac{1}{2M\hbar
\Omega}}P
  + \frac{\chi}{\hbar \Omega} x \right) \; ,
  \label{eq:operator1}\\
  L_2 &=& \sqrt{\gamma\nu}
  \left( \sqrt{\frac{M\Omega}{2\hbar}}Q-i\sqrt{\frac{1}{2M \hbar \Omega}}P
  + \frac{\chi}{\hbar \Omega} x \right)  \; ,\
 \label{eq:operator2}
\eea
where $\gamma$ is a damping parameter related to the intensity of thermal bath,
$\Omega=\sqrt{\kappa/M}$, $k_B$ is the Boltzmann constant, $T$ is temperature
and $\nu = (e^{\hbar \Omega/k_BT}-1)^{-1}$ is the Bose-Einstein distribution.

Substituting Eqs. (\ref{eq:operator1}) and (\ref{eq:operator2}) into Lindblad
master equation in Eq. (\ref{eq:lindblad}),
 \bea
 \frac{\pd \rho}{\pd t} & = &
 -\frac{i}{\hbar}[H,\rho]
 -\frac{i\delta_1}{2\hbar}[Q,P\rho+\rho P]
 \nonumber \\
 && - \frac{i \delta_2}{2\hbar}[x,P\rho+\rho P]
 -\frac{\delta_3}{2\hbar} \left[ Q,[Q,\rho] \right]
 \nonumber \\
 &&
 -\frac{\delta_4}{2\hbar} \left[x,[Q,\rho] \right]
 -\frac{\delta_5}{2\hbar} \left[x,[x,\rho] \right]
 -\frac{\delta_6}{2\hbar} \left[P,[P,\rho] \right]  \; ,\
 \label{eq:lindblad22}
\eea
where $\delta_i$'s are the coefficient in the Lindblad operators,
\bea
  \delta_1 & = & \frac{\gamma}{2 \hbar} (1 + 2 \nu) \; , \\
  \delta_2 & = & \frac{\gamma\chi}{4 \hbar \Omega} \sqrt{\frac{\omega}{\Omega}}
(1 + 2 \nu) \; , \\
  \delta_3 & = & \frac{\gamma M \Omega}{2 \hbar}(1+2 \nu) \; , \\
  \delta_4 & = & \frac{\gamma\chi}{2 \hbar \Omega} \sqrt{M m \Omega \omega}
(1+2 \nu) \; , \\
  \delta_5 & = & \frac{\gamma\chi^2 m \omega}{4 \hbar \Omega^2} (1 + 2 \nu)  \;
, \\
  \delta_6 & = & \frac{\gamma}{2 \hbar M \Omega} (1 + 2 \nu)  \; .
  \label{eq:deltanya}
\eea
$\delta_{1,2}$ are the frictional damping rate, while
$\delta_{3,4,5,6}$ are the quantum mechanical diffusion coefficients
\cite{palchikov}. This is the underlying model in the paper.

\section{Path integral calculation of partition function}
\label{sec:pi}

In order to solve the master equation in Eq. (\ref{eq:lindblad22}), the
resolution of a set of differential equations among the matrix elements of
density operator with respect to a specific basis is required. This is usually
provided by the eigen states of the system hamiltonian \cite{nakazato}. But in
the statistical mechanics it is not necessary to solve the master equation.
Instead one can calculate the partition function $\z$ using, for instance,
path integral method.

\subsection{Partition function}

The density matrix can be obtained by performing a transformation $t \rightarrow
\tau = -it = \beta \hbar$ with $\beta = 1/(k_B T)$ \cite{feynmann}.
We assume that the quantum mechanical diffusion is dominant than
the frictional damping rate such that it can be ignored. Then, the Lindblad
master equation of the unnormalized $\rho$ can be rewritten as,
\be
  -\frac{\pd \rho}{\pd \beta}= H \rho + \triangle \rho \; ,
  \label{eq:thermo1}
\ee
where
\be
  \triangle \rho =
  \frac{i\delta_3 }{2\hbar} Q^2 \rho + \frac{i\delta_4}{2\hbar} xQ\rho
+ \frac{i\delta_5}{2\hbar} x^2 \rho + \frac{i\delta_6}{2\hbar} P^2 \rho \; .
\label{eq:difusi}
\ee
It should be remarked here that, using this equation one can confirm the
Linblad operators in Eqs. (\ref{eq:operator1}) and (\ref{eq:operator2}) lead
to the right  equilibrium. The proof is given in App. \ref{app:es}.

The partition function corresponding to the master equation is given by
\cite{feynmann},
\be
 \z = \int \id [x(t)] e^{-\frac{1}{\hbar}S(x(t))} \; ,
 \label{eq:par}
\ee
where $S(x(t)) = \int_0^{\beta \hbar}(T+V) \d t'$ is the Euclidean action
corresponding to the equation.

Our interest is on the anharmonic amide-site interactions. Choosing the
amide-site potential as Eq. (\ref{eq:poten2}) and
taking into account only the first order of potential in the
interaction, the action becomes \cite{feynmann},
\bea
 \s_{xQ} & = & \int_0^\tau \d t \left(
  \frac{1}{2} m \dot{x}^2 +\frac{1}{2}m \omega^2 x^2 - \frac{1}{2}\delta x^4 +
\chi x Q
  \right. \nonumber \\
  && \left.
     + \frac{1}{2}M \dot{Q}^2 + \frac{1}{2} \kappa Q^2 - \frac{i\delta_3}{2}
Q^2- \frac{1}{4}\lambda Q^4
     \right.
  \nonumber\\
  && \left. - \frac{i \delta_4}{2} x Q
     -\frac{i\delta_5}{2}x^2 - \frac{i\delta_6}{2}\dot{Q}^2  \right) \; .
 \label{eq:inter1}
\eea
This yields,
\bea
   \s_{xQ} & = & \int_0^\tau \d t \left( \frac{1}{2} m \dot{x}^2 + \frac{1}{2}
\overline{k} x^2
         - \frac{1}{2} \delta x^4  + \tilde{\chi} x Q \right. \nonumber \\
     && \left.
       + \frac{1}{2} \overline{M} \dot{Q}^2 + \frac{1}{2} \tilde{\kappa} Q^2
        - \frac{1}{4}\lambda Q^4 \right) \; ,
\label{eq:inter3}
\eea
where $\tilde{\kappa}=\kappa-i\delta_3$ ,
$\overline{k}=k-i\delta_5$, $\overline{M}=M-i\delta_6$,
$\tilde{\chi}=\chi-i/2\delta_4$ and $k=m\omega^2$. The amide-site
is assumed to be more rigid than the amide-I. So the quantum
fluctuation is dominated by the amide-I to enable us to use the
Gaussian approximation. Making use of the Gaussian approximation,
only the classical path of $Q$ contributes to the interaction term
\cite{levit}.  Therefore the partition function becomes,
\be
  \z= \z_x \z_Q \; ,
  \label{par1}
\ee
where
\bea
   \z_x = \int \id x \, \ex^{-\s_x/\hbar} \; ,\\
   \z_Q=  \int \id Q \, \ex^{-\s_Q/\hbar} \; ,
   \label{eq:par2}
\eea
and the actions are,
\bea
   \s_{x}&=& \int_0^\tau \d t \left( \frac{1}{2}m \dot{x}^2
+\frac{1}{2}\overline{k}
x^2 - \frac{1}{2}\delta x^4
         + \tilde{\chi} x \overline{Q} \right) \; , \\
   \s_{Q}&=& \int_0^\tau \d t \left( \frac{1}{2}\overline{M} \dot{Q}^2
+\frac{1}{2}\tilde{\kappa} Q^2
   -  \frac{1}{4}\lambda Q^4  \right) \; .
 \label{eq:par3}
\eea

\subsection{Partition function for the amide-site}

Further, one should solve the partition function of the amide-site ($\z_Q$). We
use Gaussian approximation to solve the partition
function. Under this approximation the general path can be
expressed in the usual way as $Q = \overline{Q} + \breve{Q}$, where
$\overline{Q}$ is the classical path and $\breve{Q}$ is the quantum
path \cite{feynmann,levit}. Expanding the action in Taylor series
$\s_Q$ becomes \cite{levit},
\be
 \s_Q = \s_Q^\mathrm{cl} +\frac{1}{1!}\delta(\s_Q)^\mathrm{cl}
+\frac{1}{2!}\delta^2(\s_Q)^\mathrm{cl} + \cdots \; .
 \label{eq:sQ1}
\ee
Since the classical path satisfies the variational principles, $\delta\s_Q = 0$,
and taking only the second order,
\be
  \z_Q =  \int \id Q \, \exp \left[ \frac{1}{\hbar} \left( \s_Q^\mathrm{cl} +
\frac{1}{2!} \delta^2 (\s_Q^\mathrm{cl}) \right) \right] \; .
   \label{eq:sQ2}
\ee
This yields,
\bea
  \z_Q & = & \ex^{-\s_Q^\mathrm{cl}/\hbar} \int \id \breve{Q} \, \exp \left\{
    -\frac{1}{\hbar}\int \d t\frac{1}{2!} \left[ \left( \frac{\pd^2
\l^\mathrm{cl}_{\overline{Q}}}{\pd \overline{Q}^2} \right) \breve{Q}^2
   \right. \right. \nonumber \\
   && \left.\left.
    + 2 \left( \frac{\pd^2
\l^\mathrm{cl}_{\overline{Q}}}{\pd\overline{Q}\dot{\overline{Q}}} \right)
\breve{Q} \dot{\breve{Q}}
    + \left( \frac{\pd^2 \l^\mathrm{cl}_{\overline{Q}}}{\pd
\dot{\overline{Q}}^2} \right) \dot{\breve{Q}}^2
    \right] \right\} \; ,
   \label{eq:sQ3}
\eea
where,
\be
   \l^\mathrm{cl}_{\overline{Q}} = \frac{1}{2} \overline{M} \dot{\overline{Q}}^2
+ \frac{1}{2}
\tilde{\kappa} \overline{Q}^2
   -  \frac{1}{4} \lambda \overline{Q}^4  \; .\
 \label{eq:sQ4}
\ee

Calculating the second variation, and substituting the result into $\z_Q$ in Eq.
(\ref{eq:sQ3}) one has,
\bea
    \z_Q & = & \ex^{-\s_Q^\mathrm{cl}/\hbar} \, \z_{Q_0} \nonumber \\
    & = & \exp \left[ -\frac{1}{\hbar} \int_{0}^\tau \d t
    \left( \frac{1}{2} \overline{M} \dot{\overline{Q}}^2 +
\frac{1}{2}\tilde{\kappa}
\overline{Q}^2 -
    \frac{1}{4}\lambda \overline{Q}^4 \right)
    \right] \nonumber\\
   &&
   \times \int \id  \breve{Q} \, \exp \left[
    -\frac{1}{\hbar} \int_{0}^\tau \d t
   \right. \nonumber \\
   && \left.
   \times \left( \frac{1}{2} \overline{M} \dot{\breve{Q}}^2
   + \frac{1}{2}\tilde{\kappa} \breve{Q}^2 - \frac{3}{2}\lambda\overline{Q}^2
\breve{Q}^2 \right)
   \right] \; .
   \label{eq:sQ5}
\eea
The equation of motion for $\overline{Q}$ in Euclidean coordinate is given by,
\be
  \frac{d ^2\overline{Q}}{d t^2} - \overline{\Omega}^2\overline{Q} +
    \frac{\lambda}{\overline{M}} \overline{Q}^3 = 0 \; ,
    \label{eq:eomQ}
\ee
where $\overline{\Omega}^2 = {\tilde{\kappa}}/\overline{M}$. In this paper
the solution is taken to have the form of,
\be
   \overline{Q}= \overline{Q}_0 \, \mathrm{sech}(\overline{\Omega}t)\; ,
   \label{eq:solutionsQ}
\ee
which leads to $\overline{Q}_0= \overline{\Omega}\sqrt{{(2 M)}/\lambda}$. This
choice is motivated by the fact that the Davydov model is the self-trapping of
its energy, \ie it should be localized. Substituting this
solution into classical action and using the identity $1 - \mathrm{sech}^2 x =
\tanh^2 x$, one obtains,
\be
  \s_Q^\mathrm{cl}  =  A_1 \tanh \left(\overline{\Omega} \hbar \beta \right)
    + A_2 \tanh^3 \left( \overline{\Omega}\hbar\beta \right) \; ,
    \label{eq:sQklasik}
\ee where, \bea
    A_1 & = &
\frac{1}{2}\overline{M}\overline{\Omega}\overline{Q}_0^2-\frac{1}{4}\frac{
\lambda}{\overline{\Omega}
}\overline{Q}_0^4 \; , \\
    A_2 & = &
\frac{1}{12}\frac{\lambda}{\overline{\Omega}}\overline{Q}_0^4-\frac{1}{2}
\overline{M}\overline{
\Omega}\overline{Q}_0^2 \; .
    \label{eq:coefsQklasik}
\eea

Now the problem is turned into solving the prefactor in the path integral,
\bea
    \z_{Q_0} & = & \int \id  \breve{Q} \, \exp \left[ -\frac{1}{\hbar}
\int_{0}^\tau \d t
    \left( \frac{1}{2} \overline{M} \dot{\breve{Q}}^2
     \right.\right. \nonumber \\
    && \left.\left.
        + \frac{1}{2}\kappa \breve{Q}^2 - \frac{3}{2}\lambda\overline{Q}^2
    \breve{Q}^2 \right) \right] \; .
   \label{eq:prefactorsQ}
\eea
Using semi-classical approximation it can be rewritten as \cite{ranfagui},
\be
  \mathcal{Z}_{Q_0}=  \frac{1}{\sqrt{2\pi \hbar}} \left. \left(\frac{\delta^2
  \s_{\breve{Q}_0}}{\delta  \breve{Q}^2} \right)^{1/2} \right|_{\overline{Q}}
   \label{eq:prefactorsQ2}  \; ,
\ee
and its second order variation is given by,
\bea
  \left. \frac{\delta^2 \s_{\breve{Q}_0}}{\delta \breve{Q}^2}
\right|_{\overline{Q}} & = &
  \frac{\tau \sqrt{\overline{M}\tilde{\kappa} }}{2\pi \hbar} \int^\tau_0 \d t
  \left( \frac{\pd^2 \l^\mathrm{cl}_{\overline{Q}}}{\pd
\dot{\overline{Q}}^2}\ddot{\overline{Q}}^2
  \right. \nonumber \\
  && \left.
   + 2\frac{\pd^2 \l^\mathrm{cl}_{\overline{Q}}}{\pd \dot{\overline{Q}}
  \pd \overline{Q}}\overline{Q}\dot{\overline{Q}} + \frac{\pd^2
\l^\mathrm{cl}_{\overline{Q}}}{\pd \overline{Q}^2}\dot{\overline{Q}}^2 \right)
     \label{eq:prefactorsQ3}  \; .
\eea
The result is,
\be
  \left. \frac{\delta^2 \s_{\breve{Q}_0}}{\delta \breve{Q}^2}
\right|_{\overline{Q}}
=
  \frac{\tau \sqrt{\overline{M}\tilde{\kappa} }}{2\pi \hbar} \int^\tau_0 \d t
  \left[ \overline{M}\ddot{\overline{Q}}^2
    + \left( - \tilde{\kappa}  +  3 \lambda \overline{Q}^2 \right)
\dot{\overline{Q}}^2
\right]
     \label{eq:prefactors4}  \; ,
\ee
 and by substituting $\overline{Q}$ in Eq. (\ref{eq:solutionsQ}),
\bea
  \left. \frac{\delta^2 \s_{\breve{Q}_0}}{\delta \breve{Q}^2}
\right|_{\overline{Q}}
& = &
  \frac{\beta \sqrt{\overline{M}\tilde{\kappa} }}{2\pi}
   \left[ \Lambda_1 \, \mathrm{sech}^4(\overline{\Omega}\hbar \beta
)\tanh(\overline{\Omega}\hbar \beta)
   \right. \nonumber \\
  && \left.
  \times \left( \cosh(2\overline{\Omega}\hbar \beta) - 3 \right)
       + \Lambda_2 \tanh^3(\overline{\Omega}\hbar \beta )
   \right. \nonumber\\
  && \left.
  + \Lambda_3 \mathrm{sech}^2 (\overline{\Omega} \hbar\beta) \tanh^3
(\overline{\Omega}\hbar \beta )
   \right. \nonumber \\
  && \left.
  \times     (4 + \cosh(2 \overline{\Omega}\hbar \beta ) \right] \; ,
     \label{eq:prefactorsQ6}
\eea
where,
\bea
    \Lambda_1 & = & \frac{1}{2}\overline{M}\overline{Q}_0^2 \; , \\
    \Lambda_2 & = &
-\frac{1}{3}\tilde{\kappa}\overline{\Omega}\overline{Q}_0^2\; , \\
    \Lambda_3 & = & \frac{1}{5}\lambda \overline{\Omega}\overline{Q}_0^4\; .
\label{eq:coefprefactorsQ}
\eea

Finally, the complete partition function for the amide-site is
obtained, \bea
  \z_Q & = &
  \frac{1}{2\pi}\sqrt{\frac{\beta}{\hbar}}(\overline{M}\tilde{\kappa})^{1/4}
  \left\{ \Lambda_1 \, \mathrm{sech}^4(\overline{\Omega}\hbar \beta
)\tanh(\overline{\Omega}\hbar \beta)
  \right. \nonumber \\
  && \left.
  \times   (\cosh(2\overline{\Omega}\hbar \beta)-3)
   - \Lambda_2 \tanh^3(\overline{\Omega}\hbar \beta )\right. \nonumber \\
  && \left.
   + \Lambda_3 \, \mathrm{sech}^2(\overline{\Omega}
\hbar\beta)\tanh^3(\overline{\Omega}\hbar \beta )
  (4+\cosh(2\overline{\Omega}\hbar \beta))\right\}^{1/2}
   \nonumber \\
  &&
  \times \exp \left[- \frac{1}{\hbar} (A_1 \tanh(\overline{\Omega} \hbar \beta)
         + A_2 \tanh^3(\overline{\Omega}\hbar\beta)) \right] \; .
     \label{eq:partisiQ}
\eea

\subsection{Partition function for the amide-I}

For the action of amide-I, $\s_x$, the partition function is,
\bea
    \z_x & = & \int \id x \, \exp \left[ -\frac{1}{\hbar} \int \d t
\left(\frac{1}{2}m \dot{x}^2 +\frac{1}{2}\overline{k} x^2
  + \tilde{\chi} x \overline{Q}
   \right.\right.  \nonumber \\
  && \left. \left.
   - \frac{1}{2}\delta x^4 \right) \right] \; .
    \label{eq:sX}
\eea
Dividing $x$ into classical path $\overline{x}$ and quantum path $\breve{x}$,
\ie
$x=\overline{x}+\breve{x}$, and again using the Gaussian approximation, the
classical
path is,
\be
    \s_x^\mathrm{cl}= \int \d t \left( \frac{1}{2}m \dot{\overline{x}}^2
+\frac{1}{2}\overline{k} \overline{x}^2
        - \frac{1}{2}\delta \overline{x}^4 + \tilde{\chi}
\overline{x}\overline{Q} \right) \; .
      \label{eq:sXklasik1}
\ee
This can be solved by determining the classical path which is the solution of
following equation,
\be
   m\ddot{\overline{x}} - \overline{k} \overline{x} + 2\delta \overline{x}^3 =
   \tilde{\chi} \overline{Q}_0 \, \mathrm{sech} (\overline{\Omega} t ) \; .
   \label{eq:eomsXklasik}
\ee
However this is hard to be solved analytically. Instead one
can use the perturbation method, \ie calculating the solutions
order by order $\overline{x} = \overline{x}^0 + \varepsilon \overline{x}^1 +
\cdots$. Note that the inhomogeneous term is assumed being
generated from the leading order. Substituting this expansion into
Eq. (\ref{eq:eomsXklasik}) up to the leading orders one has,
\be
   m\ddot{\overline{x}}^0 - \overline{k} \overline{x}^0 + 2\delta
\overline{x}^{03} = 0 \; ,
   \label{eq:eomsXklasik0}
\ee
for the lowest order and,
\be
   m\ddot{\overline{x}}^1 - \overline{k} \overline{x}^1 + 6\delta
\overline{x}^{02}\overline{x}^{1} =
\tilde{\chi} \overline{Q}_0 \,  \mathrm{sech}( \overline{\Omega} t ) \; ,\
   \label{eq:eomsXklasik02}
\ee
for the first order. Similar to the solution in the amide-site
coordinate, since $\delta < \overline{k}$, the solution for  the zeroth
order is,
\be
   \overline{x}^0= \overline{X}_{0} \, \mathrm{sech}(\overline{\omega} t) \; ,
   \label{eq:solusisXklasik0}
\ee
where $\overline{\omega}=\overline{k}/m$,
$\overline{X}_{0}=\overline{\omega}\sqrt{m/{\delta}}$ after transforming $t
\rightarrow it$. In this solution $\delta$ must not be zero. For
the first order, the solution is $x^1=x^1_h+x^1_p$ and $x^1_h$
satisfies,
\be
   \ddot{\overline{x}}_h^1 - \overline{\omega}^2 \overline{x}_h^1 +
6\frac{\delta}{m}
\overline{X}_{0}^2 \, \mathrm{sech}^2(\overline{\omega}t)\overline{x}_h^{1} = 0
\; .
   \label{eq:eomsXklasik1h}
\ee Performing a transformation $\tau = \tanh(\overline{\omega} t)$ one
gets the associated  Legendre equation,
\be
 \frac{\d}{\d\tau} \left[ (1-\tau^2) \frac{\d\overline{x}_h^1}{\d\tau} \right]
 + \left[ l ( l + 1 ) + \frac{n^2}{1 - \tau^2} \right] \overline{x}_h^1 = 0 \; ,
 \label{eq:solhomox}
\ee where $n^2=\overline{\omega}$ and $l(l+1)= 6
{\delta}/{(m\overline{\omega})} \overline{X}_{0}^2$. The solution is $x_h^1
= x_{h_0}^1P_l^n(\tanh(\overline{\omega} t))$ where $P_l^n(\tau)$ is
the associated Legendre function. Particularly its solution
satisfies,
\be
   m\ddot{\overline{x}}_p^1 - m \overline{\omega}^2 \overline{x}_p^1
   + 6\delta \overline{X}_{0}^2 sech^2(\overline{\omega} t)\overline{x}_p^{1} =
\tilde{\chi}
\overline{Q}_0 \, \mathrm{sech}( \overline{\Omega} t ) \; .
   \label{eq:eomsxklasik1p}
\ee
The solution of this equation can be written using Green function
$G(\tau,\tau')$ as follow,
\be
  \overline{x}_p^{1} = \frac{\tilde{\chi} \overline{Q}_0}{m} \int \d\tau' \,
G(\tau,\tau')
\mathrm{sech}
  \left[ \frac{\overline{\Omega} }{\overline{\omega}}  \tanh^{-1}(\tau') \right]
\; ,\
 \label{eq:solkhususx}
\ee
and the Green function is governed by,
\bea
 && \frac{\d}{\d\tau} \left[(1-\tau^2) \frac{\d G(\tau,\tau')}{\d\tau} \right]
 + \left[ l (l + 1) - \frac{m^2}{1 - \tau^2} \right] G(\tau,\tau')
  \nonumber \\
 && = -\delta(\tau-\tau') \; .
 \label{eq:green}
\eea
The Green function is given by \cite{filho}, $G(\tau,\tau')=(-1)^n P_l^n(\tau_<)
Q_l^n(\tau_>)$
with $Q_l^n(\tau)$ is the associated Legendre functions of the second kind. Its
complete solution is,
\bea
  \overline{x} & = & \overline{X}_{0} \, \mathrm{sech}(\overline{\omega} t) +
\varepsilon
\left[ x_{h_0}^1P_l^n(\tanh(\overline{\omega} t))
    \right. \nonumber \\
  & & \left.
   + \frac{\tilde{\chi} \overline{Q}_0}{m} \int \d\tau' \, G(\tau,\tau')
\mathrm{sech}
   \left( \frac{\overline{\Omega} }{\overline{\omega}} \tanh^{-1}(\tau') \right)
\right]
\; .\
 \label{eq:completesol}
\eea
Substituting this result into Eq. (\ref{eq:sXklasik1}) one obtains the classical
action.
On the other hand, the classical action up to the first order is given by,
\bea
    \s_x^\mathrm{cl} & = & \Delta_1 \tanh(\overline{\omega} \hbar \beta) +
\Delta_2
\tanh^3(\overline{\omega} \hbar \beta)
    \nonumber\\
    && + \Delta_3 \int_0^{\beta \hbar} \d t \, \mathrm{sech}(\overline{\omega}
t) \,
\mathrm{sech}(\overline{\Omega}t)\\
    && + \varepsilon (F_1 + F_2 + F_3 + F_4 +F_5 + F_6 + F_7 + F_8)
    \nonumber    \; ,\
      \label{eq:sXklasik2}
\eea
where,
\bea
  \Delta_1 & = & \frac{1}{2}m \overline{\omega} \overline{X}_{0}^2
+\frac{\delta}{3\overline{\omega}}\overline{X}_{0}^4 \; ,\\
  \Delta_2 & = & -\frac{m}{2} \overline{\omega} \overline{X}_{0}^2 +
\frac{\delta\overline{X}_{0}^4}{6 \overline{\omega}} \; ,\\
   \Delta_3 & = & -\frac{1}{2}\tilde{\chi} \overline{X}_{0}\overline{Q}_{0} \; ,
   \label{eq:koefsXklasik2nya}
\eea
and $F_i$'s are given in App. \ref{app:1}.

Performing the same procedure as done in the previous subsection, one should
consider the prefactor of  $\mathcal{Z}_x$,
that is,
\be
  \mathcal{Z}_{x^0}=  \frac{1}{\sqrt{2\pi \hbar}} \left. \left( \frac{\delta^2
\s_{\breve{x}^0}}{\delta \breve{x}^2} \right)^{1/2} \right|_{\overline{x}}
   \label{eq:prefactorsX1}  \; ,\
\ee
where the second order variation is given by,
\bea
  \left.\frac{\delta^2 \s_{\breve{x}^0}}{\delta \breve{x}^2}
\right|_{\overline{x}} &
= &
  \frac{\beta \sqrt{m\overline{\omega}^2}}{2\pi} \int^\tau_0 \d t
  \left( \frac{\pd^2 \l^\mathrm{cl}_{\overline{x}}}{\pd
\dot{\overline{x}}^2}\ddot{\overline{x}}^2
  \right. \nonumber \\
 &&
  \left.
  + 2\frac{\pd^2 \l^\mathrm{cl}_{\overline{x}}}{\pd \dot{\overline{x}}\pd
\overline{x}}\overline{x}\dot{\overline{x}}
  + \frac{\pd^2 \l^\mathrm{cl}_{\overline{x}}}{\pd
\overline{x}^2}\dot{\overline{x}}^2 \right)
     \label{eq:prefactorsX2}  \; ,
\eea
and,
\be
\l^\mathrm{cl}_{\overline{x}} = \frac{1}{2}m \dot{\overline{x}}^2
- \frac{1}{2}m \overline{\omega}^2 \overline{x}^2+ \frac{1}{2}\delta
\overline{x}^4 + \tilde{\chi} \overline{x} \overline{Q} \; , \ee which yields,
\bea
  \left. \frac{\delta^2 \s_{\breve{x}^0}}{\delta \breve{x}^2}
\right|_{\overline{x}}
& = &
 \frac{\beta \sqrt{m\overline{\omega}^2}}{2\pi} \int^{\beta \hbar}_0 \d t \,
  \nonumber \\
 && \times \left[ m\ddot{\overline{x}}^2
   - (m\overline{\omega}^2 + 3 \delta \overline{x}^2)\dot{\overline{x}}^2
\right]
     \label{eq:prefactorsX3}  \; .\
\eea
Substituting Eq. (\ref{eq:completesol}) into Eq. (\ref{eq:prefactorsX3}) and
keeping only the first order,
\bea
   && \left.\frac{\delta^2 \s_{\breve{x}^0}}{\delta
\breve{x}^2}\right|_{\overline{x}} =
   \frac{\beta \sqrt{m\overline{\omega}^2}}{2\pi} \times
    \nonumber\\
  &&
   \left[ \Gamma_1 \, \mathrm{sech}^4(\overline{\omega}\hbar \beta)
\tanh(\overline{\omega}\hbar \beta)(\cosh(2\overline{\omega}\hbar \beta)-3)
   \right.\nonumber\\
  && \left.
    + \Gamma_2 \tanh^3(\overline{\omega} \hbar \beta)
   \right.\nonumber\\
  && \left.
    + \Gamma_3 (4+\cosh(2\overline{\omega} \hbar \beta))
\mathrm{sech}^2(\overline{\omega}
\hbar \beta) \tanh^3(\overline{\omega} \hbar \beta)
   \right.\nonumber\\
  && \left.
   + \varepsilon (G_1 + G_2 + G_3 + G_4 + G_5    \right.\nonumber\\
  && \left.+ G_6 + G_7 + G_8 +G_9 + G_{10}) \right] \; ,\
   \label{eq:prefactorsX4} \; .\
\eea
where,
\bea
   \Gamma_1 & = & -\frac{m}{2}\overline{\omega}^3\overline{X}_0^2 \; , \\
   \Gamma_2 & = & \frac{m \overline{\omega}^3 \overline{X}_0^2}{3} \; , \\
   \Gamma_3 & = & \frac{\delta \overline{\omega} \overline{X}_0^4}{5} \; ,
   \label{eq:koefprefactorX4}
\eea
and $G_i$'s are given in the  App. \ref{app:2}..

Hence, the partition function $\z_x$ is obtained,
\bea
   && \z_x = \left( \frac{\beta \sqrt{m\overline{\omega}^2}}{4\pi^2 \hbar}
\right)^{1/2} \times
   \nonumber\\
  && \left[ -\Gamma_1 \, \mathrm{sech}^4(\overline{\omega}\hbar \beta)
\tanh(\overline{\omega}\hbar \beta)(\cosh(2\overline{\omega}\hbar \beta)-3)
   \right. \nonumber\\
  && \left. + \Gamma_2 \tanh^3(\overline{\omega} \hbar \beta)
   \right. \nonumber\\
  && \left.     + \Gamma_3 (4+\cosh(2\overline{\omega} \hbar \beta)) \,
\mathrm{sech}^2(\overline{\omega} \hbar \beta)\tanh^3(\overline{\omega} \hbar
\beta)
   \right. \nonumber\\
  && \left. + \varepsilon (G_1 + G_2 + G_3 + G_4 + G_5
   \right.  \nonumber\\
  && \left.  + G_6 + G_7 + G_8 +G_9 + G_{10}) \right]^{1/2}
    \nonumber\\
  &&
    \times \exp \left\{ - \frac{1}{\hbar} \left[
           \Delta_1 \tanh(\overline{\omega} \hbar \beta)
         + \Delta_2 \tanh^3(\overline{\omega} \hbar \beta)
   \right.\right. \nonumber \\
  && \left.\left.
    + \Delta_3 \int_0^{\beta \hbar} \d t \, \mathrm{sech}(\overline{\omega} t)
\,
\mathrm{sech}(\overline{\Omega}t)
   \right. \right. \nonumber\\
  && \left.\left.  + \varepsilon (F_1 + F_2 + F_3 + F_4 +F_5 + F_6 + F_7 + F_8)
\right] \right\} \; .\
    \label{eq:partisiX}
\eea

Finally, the complete partition function density of Davydov-Scott
monomer in thermal bath, $\z = \z_x \, \z_Q$, is obtained from
Eqs. (\ref{eq:partisiQ}) and (\ref{eq:partisiX}).

\section{Thermodynamical properties of Davydov-Scott monomer}
\label{sec:tpdc}

\begin{figure}[b]
   \includegraphics[width=0.45\textwidth]{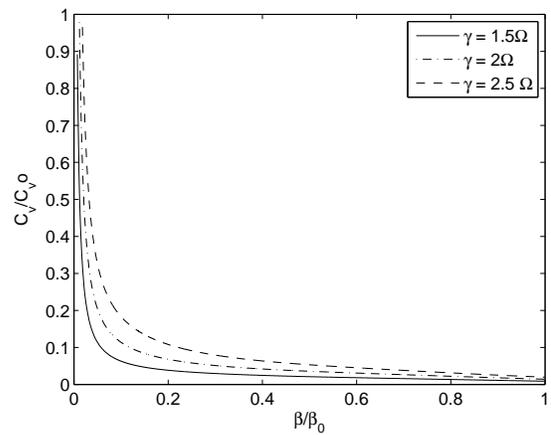}
   \caption{The temperature dependence of normalized specific heat for various
values of $\gamma$ with $\lambda = 1$.}
   \label{fig:heat}
\end{figure}

From physical point of view, the Davydov-Scott monomer is a harmonic oscillator
coupled to a quantum excitation. Using Euler-Lagrange equation, one can derive
the appropriate equation of motion (EOM) from the action in Eq.
(\ref{eq:inter1}). Although solving of the EOM is very interesting and
attractive, but it has a little physical significant due to unobservable
individual molecular motion of the Davydov-Scott monomer. Therefore in this
paper let us consider thermodynamic observable as specific heat \cite{feynmann},
\be
    C = k_B \beta^2 \frac{\pd^2 \ln (\z)}{\pd \beta^2}\; .\
     \label{eq:thermo}
\ee

Some previous works considering similar model as the Hamiltonian in
Eq. ({\ref{eq:mono1}}) to study the Davydov-Scott monomer in
thermal bath, for example the semiclassical approach in \cite{cuevas}. It has
been argued that using Lindblad formulation
the semiclassical limit is a good approximation to the corresponding full
quantum treatment at biological temperatures in the highly underdamped and
harmonic limits. In the semiclassical approximation, the coupling between
Davydov-Scott monomer with thermal bath is described by Langevin equation of the
amide-site
displacement characterized by $\gamma$ and $\Omega$ \cite{cruzeiro2}. If the
stochastic force represents the thermal bath is zero, the equation is reduced
into the damped harmonic oscillator. There are three regions regarding the
values of $\gamma$ and $\Omega$, that is $\gamma < 2\Omega$
for the underdamped condition, $\gamma = 2 \Omega$ for the critical damped
condition and $\gamma > 2 \Omega$ for the overdamped condition. In the Lindblad
formulation the damping coefficient $\gamma$ represents the relaxation time due
to interaction with the environment. The higher values of $\gamma$ corresponds
to the shorter relaxation time. The previous work by Cuevas \etal has
established that the semiclassical approximation is equivalent to the full
quantum approach (for biological temperature) as long as $\gamma \ll 2 \Omega$.
Otherwise, the oscilation frequency of the observable would be different
\cite{cuevas}.

\begin{figure}[t]
   \includegraphics[width=0.45\textwidth]{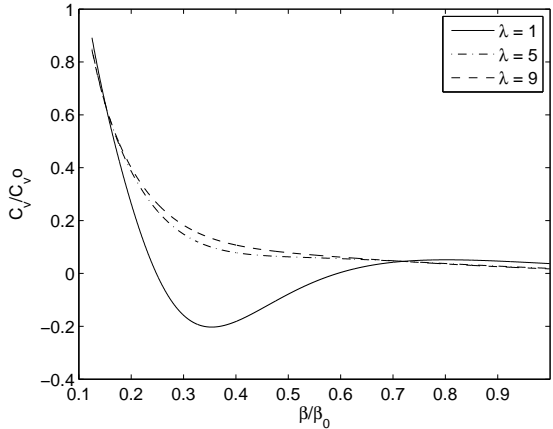}
   \caption{The temperature dependence of normalized specific heat for various
values of $\lambda$ for the underdamped condition with $\gamma = 0.1 \Omega$.}
   \label{fig:heat1}
\end{figure}

\begin{figure}[b]
   \includegraphics[width=0.45\textwidth]{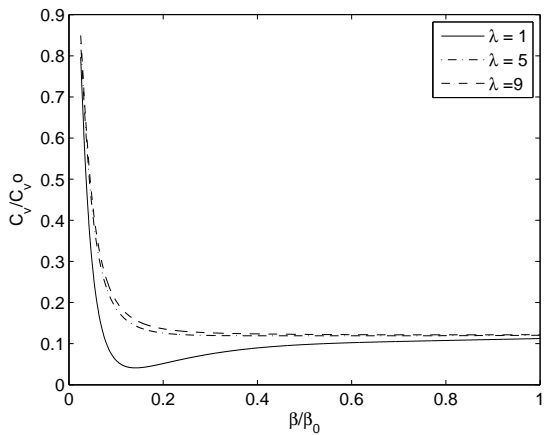}
   \caption{The temperature dependence of normalized specific heat for various
values of $\lambda$ for the critical damped condition with $\gamma = 2
\Omega$.}
   \label{fig:heat2}
 \end{figure}

\begin{figure}[t]
   \includegraphics[width=0.45\textwidth]{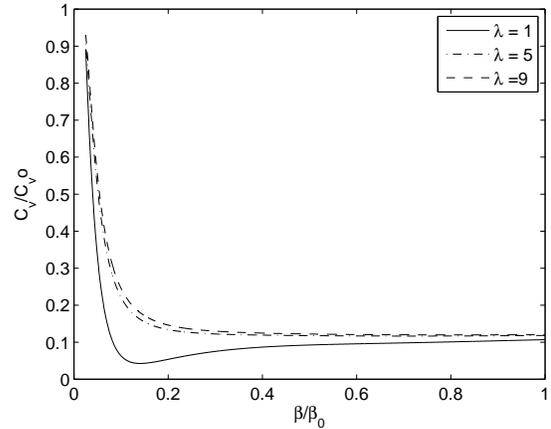}
   \caption{The temperature dependence of normalized specific heat for various
values of $\lambda$ for the overdamped condition with $\gamma = 2.5 \Omega$.}
   \label{fig:heat3}
 \end{figure}

\begin{figure}[b]
   \includegraphics[width=0.45\textwidth]{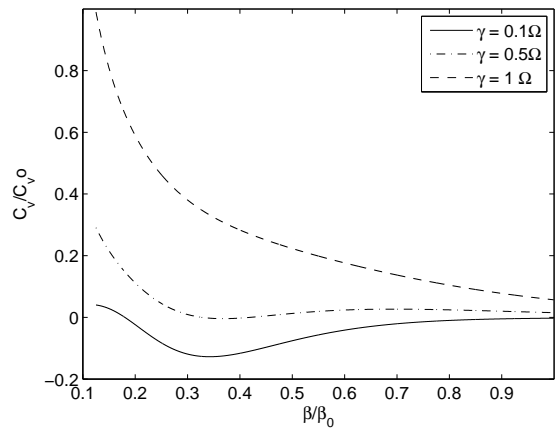}
   \caption{The temperature dependence of normalized specific heat for the
underdamped condition with various values of $\gamma < 2 \Omega$ and $\lambda =
1$.}
   \label{fig:heat4}
 \end{figure}

In this paper, the analysis is done for the above three criterions. The values
of parameters used throughout numerical calculation are $M = 6.3 \times
10^{-26}$ kg, $m = 7.3 \times 10^{-26}$ kg,
$\kappa = 10$ Nm$^{-1}$ and $\omega = 1660$ cm$^{-1}$ \cite{cuevas,sinkala},
while $\lambda = 9$. The behavior of specific heat in term of temperature is
shown in Fig. \ref{fig:heat} for various damping parameter $\gamma$, Figs.
\ref{fig:heat1}$-$\ref{fig:heat3} for various strength of anharmonic oscillation
in three damping conditions and Fig. \ref{fig:heat4} for various $\gamma$ in the
underdamped case. The results are similar with the previous ones obtained in the
calculation of a system with anharmonic amide-site \cite{zoli}.

In the present case the damping coefficients are represented by the coefficients
 $\delta_1 \sim \delta_6$. In particular, $\delta_1$ appears in the kinetic
terms, and can then be interpreted as the 'effective mass' of amide-site
vibration. Further, $\delta_3$ appears in the harmonic potential as the
'effective elastic constant'. Hence it can be argued that the
environment effects to the amide-site vibration occur through the
kinetic term and the harmonic potential. The coefficient $\delta_4$ represents
the strength of the interaction between amide-I.and the system.

From the figures, the specific heat asymptotically approaches to zero at low
temperature and to infinity at high temperature. Actually $\phi^4$ potential
oscillator interacting with electron and also the anharmonic oscillator give
similar profiles \cite{schwarcz}. Large  environment effect causes the
Davydov-Scott monomer to increase the energy and then the temperature to achieve
the equilibrium state. Correspondingly the vibration frequency is also affected
by the damping coefficients since $\omega=\sqrt{\tilde{\kappa}/\overline{M}} =
\sqrt{{(\kappa + i \delta_3)}/{(M + i \delta_2)}}$.
These results indicate that the interaction between Davydov-Scott  monomer and
thermal bath depend on the strength of the coupling of system and environment.
Recent study of the open quantum system also shows that the canonical
equilibrium state of an open quantum system depends explicitly on the
system-bath coupling strength \cite{ingold,campisi}.

In particular, from Fig. \ref{fig:heat1} the anomaly of
specific heat that becomes negative for certain parameter sets at high
temperature region is observed. The same phenomena have been pointed out by
Ingold \etal \cite{ingold,ingold2}. In the current case, the anomaly especially
appears for the underdamped condition as shown in Figs. \ref{fig:heat1} and
\ref{fig:heat4}. It is also found that the negative specific heat is restored at
large $\lambda$ and $\gamma$, \ie for large anharmonic oscillation and
intensity of thermal bath.

It should also be remarked that one cannot take $\lambda=0$ (no oscilation
effect) since all results are obtained from Eq. (\ref{eq:solutionsQ}) which is
a special case with $\lambda \ne 0$ condition.

\section{Summary}
\label{sec:su}

The interaction of Davydov-Scott monomer with thermal bath is investigated
using the Lindblad formulation of master equation. In contrast with
previous work by Cuevas \etal \cite{cuevas}, the anharmonic oscillation term of
amide-site is taken into account. Adopting similar Lindblad operators used in
\cite{cuevas}, the master equation of the system is obtained. Instead of solving
the equation of motion, the thermodynamic partition function and in particular
specific heat are calculated using the path integral methods.

It is shown that the coupling with the environment contributes to the  kinetic
term, the harmonic potential of amide-site vibration and the anharmonic term of
amide-I.

The anomaly of specific heat that becomes negative for certain parameter sets at
high temperature region is observed as pointed out  by Ingold \etal in the case
of pure open quantum systems \cite{ingold,ingold2}. However, it is found that
the negative specific heat is restored for large anharmonic oscillation effect.
In contrast to these results, Ingold \etal have found that the anomaly occurs at
low temperature region. This discrepancy can be explained as the consequences of
different approaches adopted to model the interaction between the system and the
thermal bath. Ingold \etal has represented the interaction in a set of harmonic
oscillators which becomes the coupled harmonic oscillator at classical
approximation. In contrary, in the present approach the thermal bath
is represented in a set of Lindblad operators which  becomes the underdamped
harmonic oscillator at classical approximation.

From the figures, it can in general be concluded that the anharmonic oscillation
contributes constructively to the specific heat.

\begin{acknowledgments}
AS thanks the Group for Theoretical and Computational Physics
LIPI for warm hospitality during the work. This work is funded by the Indonesia
Ministry of Research and Technology and the Riset Kompetitif LIPI in fiscal year
2010 under Contract no.  11.04/SK/KPPI/II/2010. FPZ is supported by Riset KK
2010 Institut Teknologi Bandung.
\end{acknowledgments}

\appendix

\section{The equilibrium state with the Lindblad operators in Eqs.
(\ref{eq:operator1}) and (\ref{eq:operator2})}
\label{app:es}

Let us investigate the equilibrium state in the present case. Substituting the
amide-site potential in Eq. (\ref{eq:poten2}) and the leading interaction term
in Eq. (\ref{eq:poten2}) into the Lindblad master equation in Eq.
(\ref{eq:thermo1}) yields,
\bea
   -\frac{\pd \rho}{\pd \beta} &=& \frac{1}{2m}\frac{\pd^2
   \rho}{\pd x^2} + \frac{1}{2}m\omega^2 x^2 \rho +\frac{i\delta_5}{2\hbar} x^2
\rho- \frac{1}{4}\delta
   x^4 \rho \nonumber\\
   && + \frac{1}{2M}\frac{\pd^2 \rho}{\pd Q^2} +
\frac{i\delta_6}{2\hbar}\frac{\pd^2 \rho}{\pd Q^2}  + \frac{1}{2}\kappa
   Q^2 \rho +  \frac{i\delta_3 }{2\hbar} Q^2 \rho
   \nonumber\\
   && +\frac{1}{4} \lambda Q^4\rho + \chi xQ \rho+ \frac{i\delta_4}{2\hbar} xQ
\rho \, .
   \label{eq:test1}
\eea
This can be rearranged to be,
\bea
   -\frac{\pd \rho}{\pd \beta} &=& \frac{1}{2m}\frac{\pd^2
   \rho}{\pd x^2} + \left[\frac{1}{2}m\omega^2+\frac{i\delta_5}{2\hbar}\right]
x^2 \rho - \frac{1}{4}\delta
   x^4 \rho \nonumber\\
   && + \left[\frac{1}{2M}+ \frac{i\delta_6}{2\hbar}\right]\frac{\pd^2 \rho}{\pd
Q^2}  + \left[\frac{1}{2}\kappa
   +  \frac{i\delta_3 }{2\hbar}\right] Q^2 \rho \nonumber\\
   &&  +\frac{1}{4} \lambda Q^4 \rho + \left[\chi +
\frac{i\delta_4}{2\hbar}\right] xQ \rho\, .
   \label{eq:test2}
\eea

On the other hand, the average value of any operator $A$ at a thermal
equilibrium is given by \cite{feynmann},
\be
  \langle\langle A \rangle\rangle = \frac{\mathrm{Tr}(\ex^{-\beta
H}A)}{\mathrm{Tr}(\ex^{-\beta H})} \, ,
  \label{eq:thermal1}
\ee
or in the integral form,
\be
   \langle\langle A \rangle\rangle =
   \frac{\int \rho(x) A(x) \, \d x}{\int \rho(x) \, \d x} \, .
   \label{eq:thermal2}
\ee
Before going further, it is more convenient to rewrite Eq. (\ref{eq:test2}) as
follow,
\bea
   -\frac{\pd \rho}{\pd \beta} &=& \frac{1}{2m}\frac{\pd^2
   \rho}{\pd x^2} + \frac{1}{2}\overline{m}\omega^2 x^2 \rho - \frac{1}{4}\delta
   x^4 \rho \nonumber\\
   && + \frac{1}{2\overline{M}}\frac{\pd^2 \rho}{\pd Q^2}  +
   \frac{1}{2}\overline{\kappa}  Q^2 \rho +\frac{1}{4} \lambda Q^4
   \rho +  \overline{\chi} xQ \rho \; ,
   \label{eq:pdrho1}
\eea
where $\overline{m} = m + i \delta_5/\omega^2$,
$\overline{M} = {M \hbar}/{(\hbar + i \delta_4 M)}$,
$\overline{\kappa} = \kappa + i \delta_3$ and
$\overline{\chi} = \chi + i {\delta_4}/{(2 \hbar)}$. It is clear that the
Lindblad operators shift all of the parameters, while contribute nothing to the
fourth power terms, \ie the third and sixth terms. Therefore, for the sake of
simplicity one can ignore the fourth power terms when investigating the thermal
equilibrium condition.

This fact leads to,
\bea
   -\frac{\pd \rho}{\pd \beta} & = & \frac{1}{2m}\frac{\pd^2
   \rho}{\pd x^2} + \frac{1}{2}\overline{m}\omega^2 x^2 \rho
   \nonumber\\
   && + \frac{1}{2\overline{M}}\frac{\pd^2 \rho}{\pd Q^2}  +
   \frac{1}{2}\overline{\kappa}  Q^2 \rho  +  \overline{\chi} x Q \rho \, .
   \label{eq:pdrho2}
\eea
Making use of the Gaussian approximation as before, $Q = \overline{Q} +
\breve{Q}$
\cite{feynmann}, and assuming that only the classical part of amide-I
contributes to the interaction, one immediately obtains,
\bea
   -\frac{\pd \rho}{\pd \beta} & = & \frac{1}{2m}\frac{\pd^2
   \rho}{\pd x^2} + \frac{1}{2}\overline{m}\omega^2 x^2 \rho +  \overline{\chi}
\overline{Q} x \rho
  \nonumber\\
  && + \frac{1}{2\overline{M}}\frac{\pd^2 \rho}{\pd \breve{Q}^2}  +
   \frac{1}{2}\overline{\kappa}  \breve{Q}^2 \rho \, .
   \label{eq:pdrho3}
\eea
This equation can be splitted into two equations belonging to the regular,
\be
   -\frac{\pd \rho}{\pd \beta} = \frac{1}{2\overline{M}}\frac{\pd^2 \rho}{\pd
\breve{Q}^2}  +    \frac{1}{2}\overline{\kappa}  \breve{Q}^2 \rho \, ,
   \label{eq:HObiasa}
\ee
and the driven oscillator harmonics,
\be
   -\frac{\pd \rho}{\pd \beta} =\frac{1}{2m}\frac{\pd^2
   \rho}{\pd x^2} + \frac{1}{2}\overline{m}\omega^2 x^2 \rho +  \overline{\chi}
\overline{Q} x \rho \, .
  \label{eq:drivenHO}
\ee
The solution for Eq.(\ref{eq:HObiasa}) is \cite{feynmann},
\be
  \rho=\sqrt{\frac{\overline{M}\overline{\Omega}}{2\pi \hbar \sinh(\hbar
\overline{\Omega} \beta)}}
  \exp \left[ -\frac{\overline{M}\overline{\Omega}}{\hbar} \,
  \tanh \left( \frac{1}{2} \hbar \overline{\Omega} \beta \right) \breve{Q}^2
\right] \, ,
  \label{eq:solusiHO}
\ee
with $\overline{\Omega}=\sqrt{\overline{\kappa}/\overline{M}}$. Subsequently,
the thermal equilibrium for $\breve{Q}^2$ can easily be calculated using
Gaussian integral to get,
\be
  \langle\langle \breve{Q}^2 \rangle\rangle =
  \frac{\hbar}{2\overline{M}\overline{\Omega}} \,
  \coth \left( \frac{1}{2} \hbar \overline{\Omega}  \beta \right) \, .
  \label{eq:termalQ}
\ee

The internal energy is given by,
\bea
 E &=&\frac{\hbar \overline{\Omega}}{2} \, \coth \left( \frac{1}{2}\hbar
\overline{\Omega} \beta \right)
 = \frac{\hbar \overline{\Omega}}{2}
  \frac{1 + \ex^{-\hbar \overline{\Omega} \beta}}{1 - \ex^{-\hbar
\overline{\Omega} \beta}} \nonumber\\
 & = & \frac{\hbar \overline{\Omega}}{2} + \frac{\hbar \overline{\Omega} \,
  \ex^{\hbar \overline{\Omega}\beta}}{1 - \ex^{-\hbar \overline{\Omega} \beta}}
\, .
  \label{eq:internalQ}
\eea
Meanwhile, the oscillator harmonic with $\overline{n}_Q$ amide-site has the
energy $E=\hbar \overline{\Omega} (\frac{1}{2}+\overline{n}_Q)$. Hence, the
number of quanta for amide-site at thermal equilibrium becomes,
\be
  \overline{n}_Q =\frac{\ex^{-\hbar \overline{\Omega} \beta}}{1-\ex^{-\hbar \overline{\Omega}\beta}}
    = \frac{1}{\ex^{\hbar \overline{\Omega} \beta}-1} \, ,
\label{eq:kuantaQ}
\ee
as expected. Particularly, the case of $\delta_3 = \delta_4=0$ reproduces the
oscillator harmonic at equilibrium without any environmental effects.

Following the same procedure, one can obtain the thermal equilibrium condition
for amide-I. Under the initial condition $\rho(0) = \delta(x-x')$, the
solution for Eq. (\ref{eq:drivenHO}) is \cite{feynmann},
\be
  \rho(x) = \sqrt{\frac{\overline{m}\omega}{2\pi \hbar \, \sinh(\hbar \omega
  \beta)}} \, \ex^{ \left(-A x^2 + Bx + C \right)} \, ,
 \label{eq:solusiDHO1}
\ee
where,
\bea
  A & = & \frac{1}{\hbar} \overline{m} \omega \, \tanh \left( \frac{1}{2}\hbar
\omega  \beta \right) \, ,
  \label{eq:aa}\\
  B & = & - \frac{1}{\hbar} \frac{\overline{\chi}\overline{m}\omega}{\sinh(\hbar
\omega
  \beta)} (\Lambda_1+\Lambda_2)(\ex^{\omega \hbar \beta}-1) \, ,
  \label{eq:ab}\\
    C & = & - \frac{1}{\hbar}\frac{\overline{\chi}}{4\overline{m}\omega}
\int_0^{\beta \hbar}\int_0^{\beta \hbar}
   \ex^{-\omega \mid u-u'\mid} \overline{Q}\overline{Q'} \d u \d u'\nonumber \\
   &&- \frac{1}{\hbar}\frac{\overline{\chi}\overline{m}\omega}{2\sinh(\hbar
\omega
  \beta)}
  \nonumber \\
  && \times \left[ (\Lambda_1^2+\Lambda_2^2)\ex^{\omega \hbar
\beta}-2\Lambda_1\Lambda_2 \right] \, ,
  \label{eq:ac}  \\
   \Lambda_1 &=& \frac{1}{2 \overline{m} \omega} \int_0^{\beta\hbar}
\ex^{-\omega u} \overline{Q}(u) \d u \, ,\\
  \Lambda_2 & = & \frac{1}{2\overline{m}\omega}\int_0^{\beta\hbar}
\ex^{-\omega(\beta \hbar - u)} \overline{Q}(u) \d u \, .
  \label{eq:koefDHO2}
\eea
Then the thermal equilibrium for $x^2$ is,
\be
   \langle\langle x^2 \rangle\rangle = \frac{\int \ex^{
   \left(-A x^2 + Bx + C \right)} x^2 \d x}
   {\int \ex^{ \left(-A x^2 + Bx + C \right)} \d x} \, .
  \label{eq:xtermal1}
\ee
This integral is well known, and can be calculated by performing the
transformation, $\overline{x} = B/2A$ and $\xi = x - \overline{x}$, and
defining $g(\overline{x}) = {B^2}/{4 A + C}$ as well. These yield,
\bea
   \langle\langle x^2 \rangle\rangle & = &
  \ex^{g(\overline{x})} \left[
  \int \ex^{-A\xi^2} \xi^2 \d\xi+ 2 \overline{x} \int \ex^{-A\xi^2} \xi \d\xi
    \right. \nonumber \\
  && \left. + \overline{x}^2 \int \ex^{-A\xi^2} \d\xi \right]\nonumber \\
  && \times \left[ \ex^{g(\overline{x})}\int \ex^{-A\xi^2} \d\xi \right]^{-1} \,
.
  \label{eq:xtermal2}
\eea Using the Gaussian integral, \ie $\int \ex^{-A\xi^2} \xi^2
\d\xi = \sqrt{\pi}/{(2A^{3/2})}$, $\int \ex^{-A\xi^2} \xi \d\xi=0$
and $\int \ex^{-A\xi^2} \d\xi=\sqrt{\pi}/\sqrt{A}$, the solution
is, \bea
   \langle\langle x^2 \rangle\rangle & = &  
   \frac{\displaystyle {\sqrt{\pi}}/{(2A^{3/2})} + \overline{x}^2
{\sqrt{\pi}}/{\sqrt{A}}}{\displaystyle {\sqrt{\pi}}/{\sqrt{A}}}
   \nonumber \\
  & = & \frac{1}{2A}+ \frac{B^2}{4A^2} \, .
  \label{eq:xtermal3}
\eea
Substituting Eqs. (\ref{eq:aa}) and (\ref{eq:ab}) yields,
\be
  \langle\langle x^2 \rangle\rangle =
\frac{\hbar}{2\overline{m}\omega}\coth \left( \frac{1}{2}\hbar\omega \beta
\right)
    +  \langle\langle x \overline{Q} \rangle\rangle \,  ,
 \label{eq:xtermal4}
\ee
where,
\bea
   \langle\langle x \overline{Q} \rangle\rangle  &=& \left(
\frac{\overline{\chi}}{\sinh(\hbar \omega \beta)}\right)^2
\coth^2 \left( \frac{1}{2}\hbar\omega \beta \right)
   \nonumber\\
  && \times
    \left[ (\Lambda_1+\Lambda_2)(\ex^{\omega \hbar \beta}-1) \right]^2 \, ,
   \label{eq:coplingxQ}
\eea
represents the coupling effect between amide-I and amide-site. The internal
energy is given by,
\bea
  E & = & \frac{\hbar \omega}{2} \coth \left( \frac{1}{2}\hbar\omega \beta
\right)
  + \overline{m}\omega^2 \langle\langle x \overline{Q} \rangle\rangle
  \nonumber\\
  &=& \frac{\hbar \omega}{2} + \frac{\hbar \omega \, \ex^{-\hbar
\omega\beta}}{1-\ex^{-\hbar \omega \beta}}
   + \overline{m}\omega^2 \langle\langle x \overline{Q} \rangle\rangle \, .
   \label{eq:internalx}
\eea
Again, concerning that $E=\hbar \omega (1/2 + \overline{n}_x)$, the number of
quanta for amide-I at equilibrium becomes,
\be
  \overline{n}_x = \frac{1}{\ex^{\hbar \omega \beta}-1}+\frac{\overline{m}\omega}{\hbar}
  \langle\langle x \overline{Q} \rangle\rangle \, .
 \label{eq:numberx}
\ee
The case of $\delta_3 = \delta_4=0$ reproduces the number of quanta
for amide-I at thermal equilibrium without any environmental effects.

These results confirm that the Lindblad operators defined in Eqs.
(\ref{eq:operator1}) and (\ref{eq:operator2}) lead
to the right  equilibrium as expected.

\section{The coefficients in Eq. (\ref{eq:sXklasik2})}
\label{app:1}

\bea
   F_1 & = & \frac{1}{2}m\omega x_{h_0}^1 \overline{X}_0 \, \mathrm{sech}(\omega
\hbar \beta) \, \tanh(\omega \hbar \beta)
   \nonumber\\
   && \times \left[ (-l-1+n) P_{l+1}^n \left( \tanh(\omega \hbar \beta)
   \right.\right. \nonumber \\
   && \left.\left. + P_l^n (\tanh(\omega \hbar \beta)) \right) \right] \; ,\\
   F_2 & = & -\omega x_{h_0}^1 \overline{X}_0 \, \mathrm{sech}(\omega \hbar
\beta) \, \tanh(\omega \hbar\beta)
    \nonumber\\
   && \times P_l^n(\tanh(\omega \hbar \beta)) \; , \\
   F_3 & = & \overline{X}_0 \frac{\tilde{\chi} \overline{Q}_0}{m} \left[
\mathrm{sech}(\omega \hbar \beta) - 1 \right]
    \nonumber \\
   && \times \left. \frac{\pd}{\pd \tau} \left.\left[
   \int^\tau \d\tau' \, G(\tau,\tau') \, \mathrm{sech} \left(
\frac{\overline{\Omega} }{\omega} \tanh^{-1}(\tau') \right) \right]
\right|_0^{\beta\hbar} \right. \; , \\
   F_4 & = & -\overline{X}_0 \frac{\tilde{\chi} \overline{Q}_0}{m} \,
\mathrm{sech}(\omega \hbar \beta) \, \tanh(\omega \hbar  \beta) \nonumber\\
   && \times \left. \int^{\beta\hbar}_0 \d \tau' \, G(\tau,\tau') \,
\mathrm{sech} \left( \frac{\overline{\Omega} }{\omega}    \tanh^{-1}(\tau')
\right) \right.
    \; , \\
   F_5 & = & -2\delta \overline{X}_{0}^3\overline{X}_{h_0}^1 \nonumber \\
       & & \times \int_0^{\beta \hbar} \d t \, \mathrm{sech}^3(\omega  t) \,
P_l^n(\tanh(\omega t)) \; , \\
   F_6 & = & -2 \delta \tilde{\chi} \overline{Q}_{0}\overline{X}_{0}^3
\int_0^{\beta
\hbar} \d t \, \mathrm{sech}^3(\omega t) \nonumber \\
   && \times \int^t \d\tau' \, G (\tau,\tau') \, \mathrm{sech} \left(
\frac{\overline{\Omega}}{\omega} \tanh^{-1}(\tau') \right) \; , \\
   F_7 & = & \frac{1}{2} \tilde{\chi}\overline{Q}_0\overline{X}_{h_0}^1
      \int_0^{\beta \hbar} \d t \, \mathrm{sech}(\overline{\Omega} t) \,
P_l^n(\tanh(\omega t)) \; , \\
   F_8 & = & \frac{\tilde{\chi}^2 \overline{Q}_0^2}{2m}
      \int_0^{\beta \hbar} \d t \, \mathrm{sech}(\overline{\Omega} t)
      \nonumber\\
   && \times \int^t \d\tau' \, G(\tau,\tau') \, \mathrm{sech}\left(
\frac{\overline{\Omega} }{\omega} \tanh^{-1}(\tau') \right) \; .
    \label{eq:koefsXklasik}
\eea

\section{The coefficients in Eq. (\ref{eq:prefactorsX4})}
\label{app:2}

\bea
   G_1 & = & m \overline{X}_0 \overline{X}_{h_0}^1 \omega^3 \,
\mathrm{sech}^3(\omega\hbar \beta) \left[ \cosh(2\omega\hbar \beta) - 3 \right]
\nonumber\\
   && \times \left[ -2l ( l + 1 ) + n^2 + n^2 \cosh^2(2\omega t) \right]
\nonumber\\
   && \times  P_l^n(\tanh(\omega\hbar\beta)) \; , \\
   G_2 & = & m \overline{X}_0 \omega \chi \overline{Q}_0 \,
\mathrm{sech}(\omega\hbar \beta) \, \tanh(\omega\hbar \beta)
       \nonumber\\
   && \times \left. \frac{\pd^2}{\pd \tau^2} \left.\left[\int^\tau \d\tau' \,
G(\tau,\tau') \, \mathrm{sech}\left( \frac{\overline{\Omega} }{\omega}
\tanh^{-1}(\tau') \right) \right] \right|_0^{\beta\hbar}\right. \; ,\\
   G_3 & = &\frac{m}{2}\overline{X}_0 \overline{X}_{h_0}^1\omega^3 \,
\mathrm{sech}^3(\omega\hbar \beta)
       \left[ \cosh(2\omega\hbar \beta) - 3 \right]
        \nonumber\\
   && \times \left[ (-l-1+n) \, P_{l+1}^{n}(\tanh(\omega\hbar \beta))
      \right. \nonumber\\
  && \left. + (l+1) \tanh(\omega\hbar \beta) \, P_l^n(\tanh(\omega\hbar \beta))
\right] \; , \\
   G_4 & = & \frac{1}{2} \overline{X}_0 \omega^2 \tilde{\chi} \overline{Q}_0 \,
\mathrm{sech}^3(\omega\hbar \beta)
       \left[ \cosh(2\omega\hbar \beta) - 3 \right] \nonumber\\
  && \times \left. \frac{\pd}{\pd \tau} \left.\left[ \int^\tau \d\tau' \,
G(\tau,\tau') \, \mathrm{sech}\left(\frac{\overline{\Omega} }{\omega}
\tanh^{-1}(\tau')\right) \right] \right|_0^{\beta\hbar}\right. \; , \\
   G_5 & = & -m \overline{X}_0\overline{X}_{h_0}^1\omega^4 \int_0^{\beta \hbar}
\d t \, \mathrm{sech}(\omega t) \, \tanh(\omega t)
  \nonumber\\
  && \times \left[ (-l-1+n) \, P_{l+1}^{n}(\tanh(\omega t))
   \right. \nonumber\\
  && \left. + (l+1) \tanh(\omega t) \, P_l^n(\tanh(\omega t)) \right] \; , \\
   G_6 & = & - \overline{X}_0\tilde{\chi} \overline{Q}_0\omega
       \int_0^{\beta \hbar} \d \tau \, \mathrm{sech}(\omega \tau)\tanh(\omega
\tau) \nonumber\\
   && \times \frac{\pd}{\pd \tau}\int^\tau \d\tau' \, G(\tau,\tau') \,
\mathrm{sech}\left( \frac{\overline{\Omega}}{\omega} \,  \tanh^{-1}(\tau')
\right) \; ,\\
   G_7 & = & -3 \delta\overline{X}_0\overline{X}_{h_0}^1 \omega \int_0^{\beta
\hbar} \d t
\, \mathrm{sech}^3(\omega t) \, \tanh(\omega t)
      \nonumber \\
  && \times \left[ (-l-1+n) \, P_{l+1}^{n}(\tanh(\omega t))
     \right. \nonumber\\
  && \left. + (l+1)\tanh(\omega t) \, P_l^n(\tanh(\omega t)) \right] \; ,\\
   G_8 & = & 3 \delta \overline{X}_0 \tilde{\chi} \overline{Q}_0
       \int_0^{\beta \hbar} \d\tau \, \mathrm{sech}^2(\omega \tau)
      \nonumber\\
  && \times \frac{\pd}{\pd \tau}\int^\tau \d \tau' \, G(\tau,\tau') \,
\mathrm{sech}\left( \frac{\overline{\Omega} }{\omega}  \tanh^{-1}(\tau') \right)
\; , \\
   G_9 & = & 3 \delta m \overline{X}_0^2 \overline{X}_{h_0}^1 \omega^2
\int_0^{\beta
\hbar} \d t \, \mathrm{sech}^3(\omega t) \, \tanh^2(\omega t) \nonumber\\
  && \times P_l^n(\tanh(\omega t)) \; , \\
   G_{10} & = & 3 \delta m \overline{X}_0^2 \overline{Q}_0 \omega \int_0^{\beta
\hbar}
\d\tau \, \mathrm{sech}^2(\omega \tau) \, \tanh^2(\omega \tau) \nonumber\\
  && \times \int^\tau \d \tau' \, G(\tau,\tau') \, \mathrm{sech}\left(
\frac{\overline{\Omega}}{\omega} \tanh^{-1}(\tau') \right) \; .
       \label{eq:koefprefactorsX}
\eea

\bibliography{linblad}

\end{document}